\documentstyle[amssymb,aps]{revtex}

\begin{document}
\title{Scaling of acceleration in locally isotropic turbulence}
\author{Reginald J. Hill}
\address{NOAA/Environmental Technology Laboratory, 325 Broadway, Boulder CO 80305, U.%
\\
S. A.}
\date{\today}
\maketitle

\begin{abstract}
The variances of the fluid-particle acceleration and of the
pressure-gradient and viscous force are given. \ The scaling parameters for
these variances are velocity statistics measureable with a single-wire
anemometer. \ For both high and low Reynolds numbers, asymptotic scaling
formulas are given; these agree quantitatively with DNS data. \ Thus, the
scaling can be presumed known for all Reynolds numbers. \ Fluid-particle
acceleration variance does not obey K41 scaling at any Reynolds number; this
is consistent with recent experimental data. \ The non-dimensional
pressure-gradient variance named $\lambda _{T}/\lambda _{P}$ is shown to be
obsolete.
\end{abstract}

\section{\bf Introduction}

\qquad Accelerations in turbulent flow are violent and are important to many
types of studies (La Porta {\it et al}., 2001). \ The accelerations
discussed here are caused by viscosity, $\nu \nabla _{{\bf x}}^{2}u_{i}$,
the pressure gradient, $-\partial _{x_{i}}p$, and the fluid-particle
acceleration, $a_{i}$, which are related by the Navier--Stokes equation: 
\[
a_{i}\equiv Du_{i}/Dt=-\partial _{x_{i}}p+\nu \partial _{x_{n}}\partial
_{x_{n}}u_{i}, 
\]
where $Du_{i}/Dt$ denotes the time derivative following the motion of the
fluid particle, and $u_{i}$\ is velocity. \ Here, $\nu $ is the kinematic
viscosity, $\nabla _{{\bf x}}^{2}\equiv \partial _{x_{n}}\partial _{x_{n}} $
is the Laplacian operator, $\partial $ denotes differentiation with respect
to the subscript variable, summation is implied by repeated indices,$\ p$ is
pressure divided by fluid density; density is constant. \ $\varepsilon $ is
energy dissipation rate per unit mass of fluid.

\qquad Kolmogorov's (1941) scaling (K41 scaling) uses $\varepsilon $ and $%
\nu $ as parameters, and is based on the assumption that local isotropy is
accurate for a given distance $r$ between two points of measurements, ${\bf x%
}$ and ${\bf x}^{\prime }\equiv {\bf x+r}$, $r\equiv \left| {\bf r}\right| $%
. \ Refinement of K41 to include the effects of turbulence intermittency
(Kolmogorov, 1962) leads to quantification of the Reynolds number dependence
of the deviation from K41 scaling. \ The choice of Reynolds number has been $%
R_{\lambda }\equiv u_{rms}\lambda _{T}/\nu $, based on Taylor's (1935) scale 
$\lambda _{T}\equiv u_{rms}/\left\langle \left( \partial
_{x_{1}}u_{1}\right) ^{2}\right\rangle ^{1/2}$ and $u_{rms}\equiv
\left\langle u_{1}^{2}\right\rangle ^{1/2}$, where subscript 1 denotes
components along the axis parallel to ${\bf r}$; also, from $\varepsilon
=15\nu \left\langle \left( \partial _{x_{1}}u_{1}\right) ^{2}\right\rangle $%
, $R_{\lambda }=u_{rms}^{2}/\left( \varepsilon \nu /15\right) ^{1/2}$. \
Quantities at ${\bf x}^{\prime }$\ are denoted $u_{i}^{\prime }$, $p^{\prime
}$, and $\Delta u_{i}\equiv u_{i}-u_{i}^{\prime }$, $\Delta p\equiv
p-p^{\prime }$, etc..

\qquad During 1948-1951 [Heisenberg (1948), Obukhov \& Yaglom (1951),
Batchelor (1951)], the pressure-gradient correlation, $\left\langle \partial
_{x_{i}}p\left( {\bf x},t\right) \partial _{x_{j}^{\prime }}p\left( {\bf x}%
^{\prime },t\right) \right\rangle $, mean-squared pressure gradient, $%
\left\langle \partial _{x_{i}}p\partial _{x_{i}}p\right\rangle $, and,
closely related to those, the pressure structure function, $D_{P}\left(
r\right) \equiv \left\langle \left( \Delta p\right) ^{2}\right\rangle $ were
related to the velocity structure function $D_{11}\left( r\right)
=\left\langle \left( \Delta u_{1}\right) ^{2}\right\rangle $ by means of the
assumption that $u_{i}$ and $u_{i}^{\prime }$ are joint Gaussian random
fields. \ In fact, the essential approximation, now known to be poor, is
that $\Delta u_{i}$ must be Gaussian (Hill, 1994). \ One result of that
joint Gaussian theory is that pressure-gradient acceleration has as
unobservably small effect from intermittency as does $D_{11}$. \ In fact,
pressure-gradient statistics are strongly affected by intermittency [Hill \&
Wilczak (1995, hereafter HW), Hill \& Thoroddsen (1997, hereafter HT), Hill
\& Boratav (1997, hereafter HB), Vedula \& Yeung (1999, hereafter VY), Gotoh
\& Rogallo (1999, hereafter GR), Gotoh \& Fukayama (2001, hereafter GF),
Nelkin \& Chen (1998), Antonia {\it et al}. (1999), La Porta {\it et al.}
(2001)].

\qquad A advanced theory (HW) relates $D_{P}\left( r\right) $, $\left\langle
\partial _{x_{i}}p\left( {\bf x},t\right) \partial _{x_{j}^{\prime }}p\left( 
{\bf x}^{\prime },t\right) \right\rangle $, $\left\langle \partial
_{x_{i}}p\partial _{x_{i}}p\right\rangle $, and the pressure spectrum to the
fourth-order velocity structure function: \ ${\bf D}_{ijkl}\left( {\bf r}%
\right) \equiv \left\langle \Delta u_{i}\Delta u_{j}\Delta u_{k}\Delta
u_{l}\right\rangle $. \ The theory allows calculation of such
pressure-gradient-related statistics from components of ${\bf D}_{ijkl}$\
and therefore by means of hot-wire anemometry, as in HT. \ This theory is
valid for all Reynolds numbers and is based on local isotropy without
further assumptions. \ The degree to which the theory's predictions are
accurate must depend on how anisotropic the large scales are and how large
the Reynolds number is; isotropy must be approached as Reynolds number
becomes small.\ \ A advanced theory for the correlation $\left\langle \nu
\nabla _{{\bf x}}^{2}u_{i}\nu \nabla _{{\bf x}^{\prime }}^{2}u_{i}^{\prime
}\right\rangle $ is given by HT;\ HT's relationship of this correlation to
the third-order velocity structure function ${\bf D}_{ijk}\left( {\bf r}%
\right) \equiv \left\langle \Delta u_{i}\Delta u_{j}\Delta
u_{k}\right\rangle $ has advantages (HT) over the 1948-1951 theory that
related $\left\langle \nu \nabla _{{\bf x}}^{2}u_{i}\nu \nabla _{{\bf x}%
^{\prime }}^{2}u_{i}^{\prime }\right\rangle $ to $D_{11}$. \ The advanced
theories have been used to compare $D_{P}\left( r\right) $\ calculated from $%
{\bf D}_{ijkl}\left( {\bf r}\right) $\ with $D_{P}\left( r\right) $\
calculated from DNS pressure fields (figure 1 of HB), as well as the
corresponding calculations for $\left\langle \partial _{x_{i}}p\left( {\bf x}%
,t\right) \partial _{x_{j}^{\prime }}p\left( {\bf x}^{\prime },t\right)
\right\rangle $\ and $\left\langle \nu \nabla _{{\bf x}}^{2}u_{i}\nu \nabla
_{{\bf x}^{\prime }}^{2}u_{i}^{\prime }\right\rangle $\ (figures 12, 13 of
VY). \ Since the advanced theories use only the Navier--Stokes equation,
incompressibility, and local isotropy, comparisons of data with the theory
give a measure of the local anisotropy of the data, of numerical
limitations, or of inaccuracy of Taylor's hypothesis (when used).

\qquad Most studies of turbulent acceleration use the traditional approach
of determining the $R_{\lambda }$\ dependence that results from use of K41
scaling of acceleration statistics [e.g., VY, GR, GF, Antonia {\it et al.}
(1999), La Porta {\it et al.} (2001)]. \ The resultant deviation from K41
scaling, i.e. the $R_{\lambda }$\ dependence, is often called 'anomalous'. \
There is no anomaly when the advanced theory is employed. \ Because $%
R_{\lambda }$ contains $u_{rms}$, it is affected by the large scales where
anisotropy is possible, and $R_{\lambda }$\ is therefore not a parameter of
the advanced theory. \ However, to compare the advanced theory with the
existing body of empirical knowledge, the advanced-theory's scales must be
expressed in terms of K41 scales, thereby producing dependence on $%
R_{\lambda }$ and $\varepsilon $. \ However, K41 scaling parameters and $%
R_{\lambda }$\ are not the scales within the advanced theory.

\section{\bf Scaling of mean-squared pressure gradient}

\qquad For locally isotropic turbulence, HW gave the relationship between
the mean-squared pressure gradient and the fourth-order velocity structure
function: 
\begin{equation}
\left\langle \partial _{x_{i}}p\partial _{x_{i}}p\right\rangle =\chi =4%
\stackrel{\infty }{%
\mathrel{\mathop{\int }\limits_{0}}%
}r^{-3}\left[ D_{1111}\left( r\right) +D_{\alpha \alpha \alpha \alpha
}\left( r\right) -6D_{11\beta \beta }\left( r\right) \right] dr,  \label{chi}
\end{equation}
where $\chi $ is shorthand for the mean-squared pressure gradient for the
case of local isotropy. $\ $In (\ref{chi}),$\ D_{1111}\left( r\right) $, $%
D_{\alpha \alpha \alpha \alpha }\left( r\right) $, and $D_{11\beta \beta
}\left( r\right) $ are components of ${\bf D}_{ijkl}\left( {\bf r}\right) $; 
$\alpha $ and $\beta $ denote the Cartesian axes perpendicular to ${\bf r}$,
and the $1$-axis is parallel to ${\bf r}$. \ Repeated Greek indices do not
imply summation. \ The result (\ref{chi}) applies for all Reynolds numbers
and without approximation other than local isotropy.

\qquad Defining $H_{\chi }$\ as the ratio of the integral in (\ref{chi}) to
its first term, HW wrote (\ref{chi}) as 
\begin{equation}
\chi =4H_{\chi }\stackrel{\infty }{%
\mathrel{\mathop{\int }\limits_{0}}%
}r^{-3}D_{1111}\left( r\right) dr.  \label{Hchidef}
\end{equation}
Equivalently, (\ref{Hchidef}) defines $H_{\chi }$. \ The purpose of (\ref
{Hchidef}) as stated in HW is that if the Reynolds number variation of $%
H_{\chi }$\ is known, then (\ref{Hchidef}) enables evaluation of $\chi $ by
calculating the integral in (\ref{Hchidef}) using data from a single
hot-wire anemometer. \ Further, HW argued that $H_{\chi }$\ is a constant at
large Reynolds numbers. \ VY evaluated $H_{\chi }$\ by means of DNS data and
found that it\ is constant at a value of about $0.65$ for $80<R_{\lambda
}<230$, $230$ being their maximum $R_{\lambda }$. \ Their $H_{\chi }$\ only
decreased to about $0.55$ at $R_{\lambda }=20$. \ Hill (1994) gives\ $%
H_{\chi }\rightarrow 0.36$ as $R_{\lambda }\rightarrow 0$ [on the basis that
the joint Gaussian assumption can be used in this limit and by use of a
formula for the velocity correlation for $R_{\lambda }\rightarrow 0$\ given
by Batchelor (1956)].

\qquad It is useful to express the integral in (\ref{Hchidef}) in terms of
quantities that have been measured in the past. \ On the basis of empirical
data described in Appendix A, the approximation for high Reynolds numbers is 
\begin{equation}
\chi \simeq 3.1H_{\chi }\varepsilon ^{3/2}\nu ^{-1/2}F^{0.79}\simeq
3.9H_{\chi }\varepsilon ^{3/2}\nu ^{-1/2}R_{\lambda }^{0.25}\text{ \ for \ }%
R_{\lambda }\gtrsim 400,  \label{hiRchi}
\end{equation}
where $F\equiv \left\langle \left( \partial _{x_{1}}u_{1}\right)
^{4}\right\rangle /\left\langle \left( \partial _{x_{1}}u_{1}\right)
^{2}\right\rangle ^{2}$ is a velocity-derivative flatness. \ For $H_{\chi
}=0.65$, (\ref{hiRchi}) agrees quantitatively with the DNS data in Table 1
of GF for $R_{\lambda }\geq 387$; thereby, the estimated limitation
supported in Appendix A, i.e., $R_{\lambda }\gtrsim 10^{3}$, seems too
conservative. \ That is why the limitation $R_{\lambda }\gtrsim 400$ is
given in (\ref{hiRchi}).

\qquad The case of low Reynolds numbers is in Appendix B where Taylor's
scaling and data from VY are used; the result is 
\begin{equation}
\chi \simeq 0.11\varepsilon ^{3/2}\nu ^{-1/2}R_{\lambda }\text{ \ for \ }%
R_{\lambda }\lesssim 20,  \label{loRchi}
\end{equation}
which is shown in figure 1 of VY. \ Neither (\ref{loRchi}) nor (\ref{hiRchi}%
) is K41 scaling because of their $R_{\lambda }$\ dependence.

\qquad The data of Pearson \& Antonia (2001) reveal how the approximation (%
\ref{hiRchi}) is approached as $R_{\lambda }$ increases. \ The inner scale
of $D_{1111}\left( r\right) $ is denoted by $\ell $\ and is defined in
Appendix A as the intersection of viscous- and inertial-range asymptotic
formulas for $D_{1111}\left( r\right) $. \ Thus, $\ell $\ is a length scale
in the dissipation range. \ In figures 4 and 5 of Pearson \& Antonia (2001),
the scaled components of $D_{1111}\left( r\right) $ increase most rapidly at 
$r>\ell $\ as $R_{\lambda }$\ is increased until an inertial range is
attained. \ This implies that the integral in (\ref{Hchidef}) will approach
the asymptote (\ref{hiRchi}) from below. \ That has been observed, as shown
in figure 1, wherein the DNS data of VY and GF are plotted with the
asymptotic formulas (\ref{hiRchi}) and (\ref{loRchi}); those asymptotes are
graphed to $R_{\lambda }=100$ and $40$, respectively. \ Because $H_{\chi
}=0.65$\ was used in (\ref{hiRchi}) to obtain figure 1, and because VY found 
$H_{\chi }\simeq 0.65$ for $80<R_{\lambda }<230$, it appears that $H_{\chi }$%
\ remains constant at about $0.65$\ for $R_{\lambda }>80$. \ There was no
adjustment of (\ref{hiRchi}) to cause agreement with the DNS. \ The
agreement is surprising because the empirical data used in Appendix A
suggests at least a 15\% uncertainty of the coefficients in (\ref{hiRchi}).
\ As a reminder of this fortunate circumstance, a $\pm 15\%$ error bar is
shown at $R_{\lambda }=10^{3}$\ in figure 1. \ Previously, $\left\langle
\partial _{x_{i}}p\partial _{x_{i}}p\right\rangle /\varepsilon ^{3/2}\nu
^{-1/2}\varpropto R_{\lambda }^{1/2}$ has been reported (VY, GR); this is a
good fit to the data of VY in the range $130<R_{\lambda }<300$, and $%
\left\langle \partial _{x_{i}}p\partial _{x_{i}}p\right\rangle /\varepsilon
^{3/2}\nu ^{-1/2}\varpropto R_{\lambda }^{0.62}$\ is a good fit to the data
of GR in the range $39<R_{\lambda }<170$.

\ \ \ \ \ \ \ \ \ \ \ \ \ \ \ \ \ \ \ \ \ \ \ \ \ \ {\bf SEE THE FIRST PAGE
IN ANCILLARY POSTSCRIPT FILE} 
\begin{figure}[tbp]
\caption{{\sc Figure} 1. K41 scaled mean-squared pressure gradient (sum of
squares of the 3 components). \ VY data: asterisks; GF data: triangles;
equations (2.3, 2.4): lines. \ 15\% error: vertical bar.}
\label{Figure1}
\end{figure}

In comparison with figure 1, those power laws are local fits over limited
ranges of $R_{\lambda }$; they are not asymptotic power laws. \ On the basis
of their equation (59b), HW gave the first prediction of the increase of $%
\left\langle \partial _{x_{i}}p\partial _{x_{i}}p\right\rangle /\varepsilon
^{3/2}\nu ^{-1/2}$\ with Reynolds number;\ (\ref{hiRchi}) is a refinement of
equation (59b) of HW. \ GR noted that equation (59b) gave a weaker
dependence on $R_{\lambda }$\ than their observed $\sim R_{\lambda }^{1/2}$\
dependence, found from DNS for which $R_{\lambda }<175$. \ The reason is now
apparent from figure 1: $\ R_{\lambda }<175$ is too low to use (\ref{hiRchi}%
) or equation (59b) of HW. \ Figure 1 does not support the multifractal
result (i.e., $R_{\lambda }^{0.135}$, about half the slope of (2.3)] given
by Borgas (1993).

\section{{\bf Scaling of }$\protect\nu ^{2}\left\langle \left| \protect\nabla
_{{\bf x}}^{2}u_{i}\right| ^{2}\right\rangle $}

\qquad Derivation of $\nu ^{2}\left\langle \left| \nabla _{{\bf x}%
}^{2}u_{i}\right| ^{2}\right\rangle $ from the spatial correlation of $\nu
\nabla _{{\bf x}}^{2}u_{i}$ is given by Hill (2001). \ It suffices here to
state the various equivalent formulas: 
\begin{equation}
V_{ii}\left( 0\right) =\nu ^{2}\left\langle \left| \nabla _{{\bf x}%
}^{2}u_{i}\right| ^{2}\right\rangle =12\nu ^{2}\left\langle \left( \partial
_{x_{1}}^{2}u_{\beta }\right) ^{2}\right\rangle =35\nu ^{2}\left\langle
\left( \partial _{x_{1}}^{2}u_{1}\right) ^{2}\right\rangle ,  \label{Vii00}
\end{equation}
where $\partial _{x_{1}}^{2}\equiv \partial ^{2}/\partial x_{1}^{2}$, and
for local stationarity,

\begin{eqnarray}
V_{ii}\left( 0\right) &=&\nu \left\langle \omega _{i}\omega
_{j}s_{ij}\right\rangle =-\frac{4}{3}\nu \left\langle
s_{ij}s_{jk}s_{ki}\right\rangle  \label{Vii1} \\
&=&-\frac{105}{4}\nu \left\langle \left( \partial _{x_{1}}u_{\beta }\right)
^{2}\partial _{x_{1}}u_{1}\right\rangle =-\frac{35}{2}\nu \left\langle
\left( \partial _{x_{1}}u_{1}\right) ^{3}\right\rangle =0.30\varepsilon
^{3/2}\nu ^{-1/2}\left| S\right| ,  \label{ViiS}
\end{eqnarray}
where $\omega _{i}$ is vorticity and $s_{ij}$\ is the rate of strain, and $%
S\equiv \left\langle \left( \partial _{x_{1}}u_{1}\right) ^{3}\right\rangle
/\left\langle \left( \partial _{x_{1}}u_{1}\right) ^{2}\right\rangle ^{3/2}$
is the velocity-derivative skewness. $\ \left| S\right| $\ is known
(Sreenivasan \& Antonia, 1997) to increase with increasing $R_{\lambda }$. \
Thus, all statistics in (\ref{Vii00}) and (\ref{Vii1}) have the same
increase with increasing $R_{\lambda }$\ when they are nondimensionalized
using K41 scaling. \ On the other hand, $S$ is approximately constant over
the range of about $20<R_{\lambda }<400$\ (Sreenivasan \& Antonia, 1997)
such that $V_{ii}\left( 0\right) $ approximately follows K41 scaling in that
range.

\qquad The most carefully selected data for $S$\ and $F$ at high Reynolds
numbers are those of Antonia {\it et al.} (1981), which data are in
agreement with data at $R_{\lambda }=10^{4}$\ by Kolmyansky, Tsinober \&
Yorish (2001). \ The data of Antonia {\it et al.} (1981) are used for $F$ in
(\ref{hiRchi}) and give $\left| S\right| \simeq 0.5\left( R_{\lambda
}/400\right) ^{0.11}$ for $R_{\lambda }>400$. \ Substituting $\left|
S\right| \simeq 0.5\left( R_{\lambda }/400\right) ^{0.11}$ in (\ref{ViiS})
gives 
\begin{equation}
V_{ii}\left( 0\right) \simeq 0.08R_{\lambda }{}^{0.11}\text{ \ for \ }%
R_{\lambda }>400.  \label{ViihiR}
\end{equation}
For $R_{\lambda }<20$, Tavoularis, Bennett \& Corrsin (1978), Herring \&
Kerr (1982) and Kerr (1985) show that $\left| S\right| $\ decreases and does
so more rapidly as $R_{\lambda }\rightarrow 0$. \ The data of Herring and
Kerr (1982) suggest that $\left| S\right| \simeq R_{\lambda }/5$\ for\ $%
R_{\lambda }<1$. \ Although stronger empirical evidence would be helpful, $%
\left| S\right| \simeq R_{\lambda }/5$\ will serve as the asymptotic formula
for $R_{\lambda }<1$, in which case (\ref{ViiS}) becomes 
\begin{equation}
V_{ii}\left( 0\right) \simeq 0.06\varepsilon ^{3/2}\nu ^{-1/2}R_{\lambda }%
\text{ \ for \ }R_{\lambda }<1.  \label{Herring}
\end{equation}

\section{\bf Scaling of the mean-squared fluid-particle acceleration}

\qquad For locally isotropic turbulence, the mean-squared fluid-particle
acceleration $A_{ii}\left( 0\right) $\ is $\chi +V_{ii}\left( 0\right) $\
because the correlation of $\nabla _{{\bf x}^{\prime }}^{2}u_{i}^{\prime } $%
\ and ${\bf \partial }_{x_{j}}p$\ vanishes by local isotropy (Obukhov \&
Yaglom, 1951). \ Thus, for any Reynolds number for which local isotropy is
valid 
\begin{equation}
A_{ii}\left( 0\right) =4H_{\chi }\stackrel{\infty }{%
\mathrel{\mathop{\int }\limits_{0}}%
}r^{-3}D_{1111}\left( r\right) dr-\frac{35}{2}\nu \left\langle \left(
\partial _{x_{1}}u_{1}\right) ^{3}\right\rangle .  \label{Aii0}
\end{equation}
Therefore, $A_{ii}\left( 0\right) $\ scales with the sum of two terms that
behave differently with Reynolds number.

\qquad Use of (\ref{hiRchi}) and (\ref{ViiS}, \ref{ViihiR}) in (\ref{Aii0})
gives for\ $R_{\lambda }\gtrsim 400$,

\begin{equation}
A_{ii}\left( 0\right) \simeq \varepsilon ^{3/2}\nu ^{-1/2}\left(
2.0F^{0.79}+0.3\left| S\right| \right) \simeq \varepsilon ^{3/2}\nu
^{-1/2}\left( 2.5R_{\lambda }^{0.25}+0.08R_{\lambda }^{0.11}\right) .
\label{hiRAii0}
\end{equation}
At $R_{\lambda }=400$, the term from $\chi $\ is $70$ times greater than the
term from $V_{ii}\left( 0\right) $, and $\chi $ increases much faster than $%
V_{ii}\left( 0\right) $\ for increasing $R_{\lambda }$. \ The theory of
1948-1951 used the joint Gaussian assumption and thereby greatly
underestimated $\left\langle \partial _{x_{i}}p\partial
_{x_{i}}p\right\rangle $ (HW, HT). \ That theory gives equation (3.18) of
Obukhov \& Yaglom (1951), which is the same as (\ref{hiRAii0}) with the
exception that $1.1\left| S\right| ^{-1}$ appears in place of $2.0F^{0.79}$.
\ Not only is the magnitude of $1.1\left| S\right| ^{-1}$\ smaller than $%
2.0F^{0.79}$ by a factor of $5$\ at $R_{\lambda }=400$, in addition, $\left|
S\right| ^{-1}$ decreases with further increases of $R_{\lambda }$\ contrary
to the increase of $2.0F^{0.79}$. \ An empirical result that seems accurate
for a variety of flows for $R_{\lambda }\gtrsim 400$\ (Champagne, 1978;
Antonia {\it et al.}, 1981) is $\left| S\right| =0.25F^{3/8}$, such that (%
\ref{hiRAii0}) can be written as $A_{ii}\left( 0\right) \simeq \varepsilon
^{3/2}\nu ^{-1/2}\left( 2.0F^{0.79}+0.075F^{0.375}\right) $ for $R_{\lambda
}\gtrsim 400$.

\qquad For low Reynolds numbers,\ use of (\ref{loRchi}) in (\ref{Aii0})
gives 
\begin{equation}
A_{ii}\left( 0\right) \simeq \varepsilon ^{3/2}\nu ^{-1/2}\left(
0.11R_{\lambda }+0.3\left| S\right| \right) \text{\ \ for \ }R_{\lambda }<20.
\label{loRAii0}
\end{equation}
From figure 8 of Herring \& Kerr (1982) one sees that $0.11R_{\lambda
}>0.3\left| S\right| $\ even at their minimum $R_{\lambda }$\ of $0.5$, and
that $0.11R_{\lambda }$ increases rapidly relative to $0.3\left| S\right| $\
as $R_{\lambda }$\ increases. \ Thus, (\ref{loRAii0}) shows that the term
from $\chi $ is the larger contribution to $A_{ii}\left( 0\right) $\ for all 
$R_{\lambda }$. \ The behaviour of $A_{ii}\left( 0\right) $, $\chi $, and $%
V_{ii}\left( 0\right) $ for moderate $R_{\lambda }$\ is shown particularly
well in figure 1 of VY.

\section{\bf Discussion}

\subsection{{\bf Obsolescence of }$\protect\lambda _{T}/\protect\lambda _{P}$%
}

\qquad The length scales $\lambda _{P}\equiv u_{rms}^{2}/\left\langle \left(
\partial _{x_{1}}p\right) ^{2}\right\rangle ^{1/2}$ and $\lambda _{T}\equiv
u_{rms}/\left\langle \left( \partial _{x_{1}}u_{1}\right) ^{2}\right\rangle
^{1/2}$ were introduced by Taylor (1935) (he included a factor of $\sqrt{2}$%
\ that has historically been dropped from these definitions), and the ratio $%
\lambda _{T}/\lambda _{P}$, and he gave the first evaluation of $\lambda
_{T}/\lambda _{P}$\ from turbulent diffusion measurements. $\ \left( \lambda
_{T}/\lambda _{P}\right) ^{2}=\left\langle \left( \partial _{x_{1}}p\right)
^{2}\right\rangle /\left[ u_{rms}^{2}\left\langle \left( \partial
_{x_{1}}u_{1}\right) ^{2}\right\rangle \right] $\ is a scaled mean-squared
pressure gradient; that scaling depends on a large-scale parameter, $u_{rms}$%
. \ As such $\lambda _{T}/\lambda _{P}$ is not relevant in the advanced
theory except in the limit of $R_{\lambda }\rightarrow 0$. \ Batchelor
(1951) obtained, for very large Reynolds numbers a dependence of $\lambda
_{T}/\lambda _{P}\varpropto R_{\lambda }^{-1/2}$, which is not correct
because it is based on the joint Gaussian assumption. \ By attempting
evaluation of $\lambda _{T}/\lambda _{P}$, Batchelor (1951) was, in effect,
attempting to enable determination of $\chi $\ from measurements of velocity
variance and energy dissipation rate. \ The advanced theory replaces $\left(
\lambda _{T}/\lambda _{P}\right) ^{2}$ with $H_{\chi }$. \ Evaluation of $%
H_{\chi }$\ allows $\chi $\ to be determined from measurement of a single
velocity component and the simple formula (\ref{Hchidef}). \ Figure 3 of GR
shows $\lambda _{T}/\lambda _{P}$ versus $R_{\lambda }$\ and reveals the
following: i) the strong $R_{\lambda }$\ dependence of $\lambda _{T}/\lambda
_{P}$\ that $H_{\chi }$\ does not have; ii) the deviation of the $R_{\lambda
}$\ dependence of $\lambda _{T}/\lambda _{P}$\ from that predicted by the
joint Gaussian assumption; (this was also found by VY) and iii) $R_{\lambda
}<20$ is required to approach the low-Reynolds-number asymptote. \ Whereas $%
H_{\chi }$ depends only on the small scales of turbulence, the dependence of 
$\lambda _{T}/\lambda _{P}$ on the large scales via $u_{rms}$\ shows that $%
\lambda _{T}/\lambda _{P}$ is not relevant in the advanced theory of local
isotropy, except in the limit of $R_{\lambda }\rightarrow 0$; for that limit
HW shows that $H_{\chi }\varpropto \left( \lambda _{T}/\lambda _{P}\right)
^{2}$.

\subsection{Acceleration data}

\qquad Pioneering technology for measuring turbulence accelerations is being
developed at Cornell. \ La Porta {\it et al.} (2001) report fluid-particle
acceleration measured in a cylindrical enclosure containing turbulent water
driven by counter-rotating blades. \ Here, `$x$' and `$y$' axes are
transverse and parallel to their cylinder axis, respectively. \ The
acceleration's flatness factor $\left\langle a_{x}^{4}\right\rangle
/\left\langle a_{x}^{2}\right\rangle ^{2}$\ in their figure 3 reaches a
maximum at $R_{\lambda }\approx 700$, and is decreased at their next-higher $%
R_{\lambda }$\ value, namely 970, the same is true for their K41-scaled
acceleration variances $\left\langle a_{i}^{2}\right\rangle /\varepsilon
^{3/2}\nu ^{-1/2}$ in their figure 4, where $i=x$ and $y$ ). \ From (\ref
{hiRAii0}), $A_{ii}\left( 0\right) /\varepsilon ^{3/2}\nu
^{-1/2}=3\left\langle a_{i}^{2}\right\rangle /\varepsilon ^{3/2}\nu ^{-1/2}$
is monotonic with $R_{\lambda }$, unlike in figure 4 of La Porta {\it et al.}%
\ (2001). \ This suggests that the cause of the maxima in flatness and
variance have the same cause.\ \ Their estimates of $R_{\lambda }$\ and $%
\varepsilon $\ both depend on the choice of a velocity component; because
the turbulence is anisotropic, the choice of another velocity component will
shift their data points along both ordinate and abscissa.\ \ This
nonuniversality of their scaling is implicated by the disappearance of the
maximum in the flatness of $\partial _{x_{1}}u_{1}$\ as presented by Belin 
{\it et al.} (1997) when $R_{\lambda }$\ is replaced by a universal Reynolds
number (Hill, 2001). \ Belin {\it et al. (}1997{\it ) }measure near one
counter-rotating blade whereas LaPorta {\it et al.} (2001) measure in the
flow's center. \ The two cylinders have different aspect ratios.{\it \ \ }It
is nevertheless instructive to substitute the values of $F$ and $\left|
S\right| $ measured by Belin {\it et al.} (1997) into (\ref{hiRAii0}),
divide by 3 to obtain the variance of one component of acceleration, and
compare the result with the data of La Porta {\it et al. (}2001{\it )}. \
This is done in figure 2, wherein the data of VY and GF are shown to agree
with the data of Belin {\it et al.} (1997). \ For the Belin {\it et al.}
data, (\ref{hiRAii0}) produces a maximum similar to that of the La Porta 
{\it et al. (}2001{\it )} data.

\ \ \ \ \ \ \ \ \ \ \ \ \ \ \ \ \ \ \ \ \ \ \ \ \ \ {\bf SEE THE SECOND PAGE
IN ANCILLARY POSTSCRIPT FILE} 
\begin{figure}[tbp]
\caption{{\sc Figure} 2 K41 scaled mean-squared acceleration component. \ La
Porta {\it et al.} (2001) data, x-component: asterisks; y-component:
squares; Belin{\it \ et al.} (1997) data in equation (\ref{hiRAii0}):
diamonds; combined DNS data of VY and GF: triangles; eq.(4.3) divided by 3:
solid line.}
\label{Figure2}
\end{figure}

As Belin {\it et al. (}1997{\it )} point out, the maximum in their data for $%
F$\ might be specific to the flow between counter-rotating blades. \ If so,
the same is likely true of the data of La Porta {\it et al.} (2001) such
that their data do not support K41 scaling of acceleration, and therefore do
not contradict the advanced theory. \ Another possibility is that the data
of LaPorta et al. (2001) at $R_{\lambda }=970$ is underestimated for unknown
reasons. \ The conclusion suggested by figure 2 and the above uncertainties
in interpretation of the data is that the data supports the scaling given
here and that such important acceleration measurements must continue.

\section{\bf Conclusion}

\qquad The asymptotes (\ref{hiRchi}) and (\ref{loRchi}) combined with the
DNS data in figure 1 determine $\left\langle \partial _{x_{i}}p\partial
_{x_{i}}p\right\rangle $ for all Reynolds numbers. For all Reynolds numbers
the advanced-theory scaling is that $\left\langle \partial _{x_{i}}p\partial
_{x_{i}}p\right\rangle $\ scales with the integral in (\ref{chi}). \ For\ $%
R_{\lambda }<20$, $\left\langle \partial _{x_{i}}p\partial
_{x_{i}}p\right\rangle $ scales with $\varepsilon ^{3/2}\nu
^{-1/2}R_{\lambda }$. \ Because $H_{\chi }$\ is apparently constant for $%
R_{\lambda }>80$, $\left\langle \partial _{x_{i}}p\partial
_{x_{i}}p\right\rangle $\ scales with $\stackrel{\infty }{%
\mathrel{\mathop{\int }\limits_{0}}%
}r^{-3}D_{1111}\left( r\right) dr$ [i.e., the integral in (\ref{Hchidef})],
and $\left\langle \partial _{x_{i}}p\partial _{x_{i}}p\right\rangle $ scales
approximately with $\varepsilon ^{3/2}\nu ^{-1/2}F^{0.79}$for $R_{\lambda
}>400$, and $\varepsilon ^{3/2}\nu ^{-1/2}F^{0.79}$\ is a good approximation
when $R_{\lambda }$ as small as $200$ (see figure 1). \ Given that $H_{\chi
} $\ is constant for $R_{\lambda }>80$, $\left\langle \partial
_{x_{i}}p\partial _{x_{i}}p\right\rangle $ could be obtained for $R_{\lambda
}>80$ using data from a single-wire anemometer by evaluating the integral in
(\ref{Hchidef}); DNS is not necessary. Using velocity data, it is more
accurate to evaluate the integral in (\ref{Hchidef}) than its approximation $%
\varepsilon ^{3/2}\nu ^{-1/2}F^{0.79}$ because evaluating $F$\ requires
greater spatial resolution than does evaluation of the integral for the same
level of accuracy. \ Evaluating $H_{\chi }$\ from (\ref{Hchidef})\ using DNS
at $R_{\lambda }>230$\ would be useful.

\qquad Now, $V_{ii}\left( 0\right) $, does scale with any of the derivative
moments in (\ref{Vii00}). \ It does not scale as in the K41 prediction
(i.e., $\varepsilon ^{3/2}\nu ^{-1/2}$) except for those $R_{\lambda }$\ at
which $S$\ is constant. \ The statement in VY that $V_{ii}\left( 0\right) $\
does obey K41 scaling is based on their data, which are within the $%
R_{\lambda }$ range where $S$\ is constant.

\qquad Fluid-particle acceleration variance, $A_{ii}\left( 0\right) $, does
not scale as in the K41 prediction (i.e., $\varepsilon ^{3/2}\nu ^{-1/2}$)
at large Reynolds numbers because of the factor $\left( 2.5R_{\lambda
}^{0.25}+0.08R_{\lambda }^{0.11}\right) $ in (\ref{hiRAii0}). $\
A_{ii}\left( 0\right) $ does not approach K41 scaling as $R_{\lambda
}\rightarrow 0$ because (\ref{Herring}) and (\ref{loRAii0}) give $%
A_{ii}\left( 0\right) \simeq 0.17\varepsilon ^{3/2}\nu ^{-1/2}R_{\lambda }$\
for $R_{\lambda }<1$. \ For all Reynolds numbers, fluid particle
acceleration does scale with the sum of velocity statistics that appears on
the right-hand side of (\ref{Aii0}).

\qquad The advanced theory is devoid of statistics of the large scales. \ It
seems paradoxical that $R_{\lambda }$, which depends on the large scales
through $u_{rms}$, is used above to delineate asymptotic regimes. \ However,
an alternative Reynolds number that depends only on small scales (Hill,
2001) makes the advanced theory self-contained. \ Use of existing
phenomenology caused both $\varepsilon $\ and $R_{\lambda }$ to appear in
this paper.\ \ However, practical applications result. \ Turbulent
acceleration-induced coalescence of droplets might be key to understanding
rain initiation from liquid-water clouds (Shaw \& Oncley, 2001). \ Radars
can measure $u_{rms}$ and $\varepsilon $, then $R_{\lambda
}=u_{rms}^{2}/\left( \varepsilon \nu /15\right) ^{1/2}$ can be determined;
then the three acceleration variances can be determined from equations given
here. \ The present results thereby support radar remote sensing of clouds
and cloud microphysical research.

\bigskip

{\bf Appendix A: High-Reynolds-number asymptote}

\qquad The lognormal model of Kolmogorov (1962) is used here; the result is
found to be insensitive to the intermittency model used. \ The inertial
range formulas are: \ $D_{1111}\left( r\right) =C^{\prime }\varepsilon
^{4/3}r^{q}L^{2\mu /9}$, $q=\left( 4/3\right) -\left( 2\mu /9\right) $, and $%
D_{11}\left( r\right) =C\varepsilon ^{2/3}r^{p}L^{-\mu /9}$, $p=\left(
2/3\right) +\left( \mu /9\right) $; $L$\ is the integral scale; $\mu =0.25$
is used (Sreenivasan \& Kailasnath, 1993), as is $C=2$ (Sreenivasan, 1995).
\ Viscous-range formulas are used; they are: $D_{1111}\left( r\right)
=\left\langle \left( \partial _{x_{1}}u_{1}\right) ^{4}\right\rangle r^{4}$\
and $D_{11}\left( r\right) =\left\langle \left( \partial
_{x_{1}}u_{1}\right) ^{2}\right\rangle r^{2}=\left( \varepsilon /15\nu
\right) r^{2}$. \ The inner scale of $D_{1111}\left( r\right) $, named $\ell 
$, is defined by equating the inertial-range formula with the viscous-range
formulas at $r=\ell $. \ In the integrand in (\ref{Hchidef}), $%
D_{1111}\left( r\right) $ can be scaled by $D_{1111}\left( \ell \right) $\
and \ $r$ by $\ell $. \ Doing so, the integral equals $\left( 3/2\right)
\ell ^{2}\left\langle \left( \partial _{x_{1}}u_{1}\right) ^{4}\right\rangle 
$ [HW showed that the remaining dimensionless integral has a value of $3/2$
for large Reynolds numbers; this is based on use of an equation for $%
D_{1111}\left( r\right) $ that is the same as equation (12) by Stolovitzky,
Sreenivasan \& Juneja (1993), who demonstrate its empirical basis]. \ Next,
the otherwise irrelevant Taylor's scale $\lambda _{T}$\ is introduced to
make use of published empirical data. \ The definition of $\ell $ and the
inertial range formulas are used to obtain $\left( \ell /\eta \right)
^{4-q}=\left( 15C\right) ^{2}\left( \lambda _{T}^{2}/\eta L\right) ^{2\mu
/9}F\left( \lambda _{T}\right) /F$; $F\left( \lambda _{T}\right) \equiv
D_{1111}\left( \lambda _{T}\right) /\left[ D_{11}\left( \lambda _{T}\right) %
\right] ^{2}$; $F\equiv \left\langle \left( \partial _{x_{1}}u_{1}\right)
^{4}\right\rangle /\left\langle \left( \partial _{x_{1}}u_{1}\right)
^{2}\right\rangle ^{2}$. \ The essential approximation is that $\lambda _{T}$%
\ is in the inertial range (Antonia {\it et al.} 1982; Pearson \& Antonia,
2001).\ \ Then, $\left( 3/2\right) \ell ^{2}\left\langle \left( \partial
_{x_{1}}u_{1}\right) ^{4}\right\rangle =\left( 3/2\right) C^{4/\left(
4-q\right) }15^{2\left( q-2\right) /\left( 4-q\right) }\left[ \left( \lambda
_{T}^{2}/\eta L\right) ^{2\mu /9}F\left( \lambda _{T}\right) /F\right]
^{2/\left( 4-q\right) }F\varepsilon ^{3/2}\nu ^{-1/2}$.\ \ For $%
10^{2}<R_{\lambda }<5\times 10^{3}$, Zocchi {\it et al.} (1994) have $%
\lambda _{T}^{2}/L\eta =30R_{\lambda }^{-1/2}$ such that $\left( \lambda
_{T}^{2}/L\eta \right) ^{\left( 2\mu /9\right) \left( 2/\left( 4-q\right)
\right) }=\left( R_{\lambda }/900\right) ^{-2\mu /\left[ 9\left( 4-q\right) %
\right] }$; the exponent of $\lambda _{T}^{2}/L\eta $ is about $0.04$; so $%
R_{\lambda }$ can vary greatly but $\left( \lambda _{T}^{2}/L\eta \right)
^{0.04}$\ remains near unity.\ \ The most consistent data for $F$ at high
Reynolds numbers are those of Antonia {\it et al.} (1981), which give $%
F\simeq 1.36R_{\lambda }^{0.31}$; the same data (Antonia {\it et al.,} 1982)
give $F\left( \lambda _{T}\right) \varpropto R_{\lambda }^{\mu }$ and $%
F\left( \lambda _{T}\right) \simeq 7.1$\ at $R_{\lambda }=9400$ so $F\left(
\lambda _{T}\right) \simeq 0.72R_{\lambda }^{\mu }$.\ \ Thus, $\left(
3/2\right) \ell ^{2}\left\langle \left( \partial _{x_{1}}u_{1}\right)
^{4}\right\rangle \simeq 0.714R_{\lambda }^{-0.064}F\varepsilon ^{3/2}\nu
^{-1/2}=0.76F^{0.79}\varepsilon ^{3/2}\nu ^{-1/2}=0.97R_{\lambda
}^{0.25}\varepsilon ^{3/2}\nu ^{-1/2}$, i.e.,

\begin{equation}
\stackrel{\infty }{%
\mathrel{\mathop{\int }\limits_{0}}%
}r^{-3}D_{1111}\left( r\right) dr\simeq 0.76F^{0.79}\varepsilon ^{3/2}\nu
^{-1/2}\simeq 0.97R_{\lambda }^{0.25}\varepsilon ^{3/2}\nu ^{-1/2}. 
\eqnum{A1}  \label{integral1}
\end{equation}
Given that $\mu \approx 0.25$, (\ref{integral1}) is insensitive of the value
of $\mu $ and is therefore also insensitive to the choice of intermittency
model. \ Shaw \& Oncley (2001) used data from the atmospheric surface layer
at $R_{\lambda }=1500$\ to obtain that (\ref{integral1}) balances to within
the accuracy of their value of $F$, i.e., about 15\%.

\qquad The data of Pearson \& Antonia (2001) show $D_{\beta \beta \beta
\beta }\left( \lambda _{T}\right) /D_{1111}\left( \lambda _{T}\right) $
becoming constant as $R_{\lambda }\rightarrow 10^{3}$\ from below; such a
constant value is a reasonable criterion for the integral in (\ref{chi}) to
be proportional to $\stackrel{\infty }{%
\mathrel{\mathop{\int }\limits_{0}}%
}r^{-3}D_{1111}\left( r\right) dr$ as in (\ref{Hchidef}). \ They show the
variation of $D_{1111}\left( r\right) $ and $D_{\beta \beta \beta \beta
}\left( r\right) $ as $R_{\lambda }$\ varies from 38\ to 1200 such that an
inertial range appears at the larger $R_{\lambda }$. \ The above
approximation (\ref{integral1}) requires that the integral in (\ref
{integral1}) approximately converges at its upper limit within the inertial
range, for which a reasonable criterion is that there be about one decade of
the power-law. \ The data of Pearson \& Antonia (2001) show such an extent
of the power law as $R_{\lambda }\simeq 10^{3}$\ is attained. \ Thus, $%
R_{\lambda }\simeq 10^{3}$\ is a well-supported lower bound for the
high-Reynolds-number approximation (\ref{integral1}).

{\bf Appendix B: \ Low-Reynolds-number asymptote}

\qquad The joint Gaussian assumption is not used here. \ Taylor's (Taylor,
1935) scaling is used; i.e., scales $\lambda _{T}$\ and $u_{rms}$\ are used
when $R_{\lambda }\rightarrow 0$. \ Taylor's scaling gives $D_{1111}\left(
r\right) \varpropto u_{rms}^{4}$, $H_{\chi }$ approaches a constant as $%
R_{\lambda }\rightarrow 0$. \ Let $x=r/\lambda _{T}$. \ Then (\ref{Hchidef})
can be written as $\chi \varpropto u_{rms}^{4}\lambda _{T}^{-2}\stackrel{%
\infty }{%
\mathrel{\mathop{\int }\limits_{0}}%
}x^{-3}D_{1111}\left( x\right) /u_{rms}^{4}dx$. \ The dimensionless integral
is a number, so $\chi \varpropto u_{rms}^{4}\lambda _{T}^{-2}$; hence $\chi
/\left( \varepsilon ^{3/2}\nu ^{-1/2}\right) \varpropto R_{\lambda }$. \
This behaviour is shown in figure 1 of VY where it appears to become
accurate between $R_{\lambda }=20$ and $40 $. \ In their Table II, $\chi
/\left( 3\varepsilon ^{3/2}\nu ^{-1/2}\right) =0.74$ at $R_{\lambda }=21$. \
Thus, $\chi =0.106\varepsilon ^{3/2}\nu ^{-1/2}R_{\lambda }$ for $R_{\lambda
}<20$. \ Compare $0.106$\ with the prediction of the joint Gaussian
assumption: $6/15^{3/2}=0.103$ (Hill, 1994).

\ \ \ \ \ \ \ \ \ \ \ \ \ \ \ \ \ \ \ \ \ \ \ \ \ \ \ \ \ \ \ \ \ \ \ \
REFERENCES

{\sc Antonia, R. A., Bisset, D. K., Orlandi, P. \& Pearson, B. R.} 1999
Reynolds number dependence of \ the second-order turbulent pressure
structure function. {\it Phys. Fluids} {\bf 11}, 241-243.

{\sc Antonia, R. A., Chambers, A. J. \& Satyaprakash, B. R.} 1981 Reynolds
number dependence of high-order moments of the streamwise turbulent velocity
derivative. {\it Bound.-Layer Meteorol.} {\bf 21}, 159-171.

{\sc Antonia, R. A., Satyaprakash, B. R. \& Chambers, A. J.} 1982 Reynolds
number dependence of velocity structure functions in turbulent shear flows. 
{\it Phys. Fluids} {\bf 25}, 29-37.

{\sc Batchelor, G. K.} 1951\ Pressure fluctuations in isotropic turbulence. 
{\it Proc. Camb. Philos. Soc.} {\bf 47}, 359-374.

{\sc Batchelor, G. K.} 1956\ {\it The Theory of Homogeneous Turbulence}.
Cambridge Univ. Press.

{\sc Belin, F., Maurer, J., Tabeling, P. \& Willaime, H.}\ 1997\ Velocity
gradient distributions in fully developed turbulence: An experimental study. 
{\it Phys. Fluids} {\bf 9}, 3843-3850.

{\sc Borgas, M. S.} 1993\ The multifractal Lagrangian nature of turbulence. 
{\it Phil. Trans. Roy. Soc. Lond. A.} {\bf 342}, 379-411.

{\sc Champagne, F. H.} 1978 The fine-scale structure of the turbulent
velocity field. {\it J.~Fluid Mech.} {\bf 86}, 67-108.

{\sc Gotoh, T. \& Fukayama, D.} 2001\ Pressure spectrum in homogeneous
turbulence. {\it Phys. Rev. Lett.} {\bf 86},\ 3775-3778.

{\sc Gotoh, T. \& Rogallo, R. S.} 1999\ Intermittency and scaling of
pressure at small scales in forced isotropic turbulence. {\it J.~Fluid Mech.}
{\bf 396}, 257-285.

{\sc Heisenberg, W.} 1948 Zur statistichen theorie der turbulenz. {\it %
A.~Physik} {\bf 124}, 628-657.

{\sc Herring, J. R. \& Kerr, R. M.} 1982\ Comparison of direct numerical
simulations with predictions of two-point closures for isotropic turbulence
convecting a passive scalar. {\it J.~Fluid Mech.} {\bf 118}, 205-219.

{\sc Hill, R. J.} 1994 The assumption of joint Gaussian velocities as
applied to the pressure structure function. {\it NOAA Tech. Rept.} ERL 451-
ETL-277 (www.bldrdoc.gov/library).

{\sc Hill, R. J.} 2001 Alternative to $R_{\lambda }$-scaling of small-scale
turbulence statistics. xxx.lanl.gov.physics/0102056.

{\sc Hill, R. J. \& Boratav, O. N.} 1997 Pressure statistics for locally
isotropic turbulence. {\it Phys. Rev. E} {\bf 56}, R2363-R2366.

{\sc Hill, R. J. \& Thoroddsen S. T.} 1997 Experimental evaluation of
acceleration correlations for locally isotropic turbulence. {\it Phys. Rev. E%
} {\bf 55}, 1600-1606.

{\sc Hill, R. J. \& Wilczak, J. M.} 1995 Pressure structure functions and
spectra for locally isotropic turbulence. {\it J.~Fluid Mech.} {\bf 296},
247-269.

{\sc Kerr, R. M.} 1985\ Higher-order derivative correlations and the
alignment of small-scale structures in isotropic numerical turbulence. {\it %
J.~Fluid Mech.} {\bf 153}, 31-58.

{\sc Kolmogorov, A. N.} 1941 The local structure of turbulence in
incompressible viscous fluid for very large Reynolds numbers. {\it Dokl.
Akad. Nauk SSSR} {\bf 30}, 538-540.

{\sc Kolmogorov, A. N.} 1962\ A refinement of previous hypotheses concerning
the local structure of turbulence in a viscous incompressible fluid at high
Reynolds number. {\it J.~Fluid Mech.} {\bf 13}, 82-85.

{\sc Kolmyansky, M., Tsinober, A. \& Yorish, S.}\ 2001\ Velocity derivatives
in the atmospheric surface layer at Re$_{\lambda }=10^{4}$. {\it Phys. Fluids%
} {\bf 13}, 311-314.

{\sc La Porta, A., Voth, G. A., Crawford, A. M., Alexander, A. \&
Bodenschatz, E.}\ 2001\ Fluid particle accelerations in fully developed
turbulence. {\it Nature} {\bf 409}, 1017-1019.

{\sc Nelkin, M. \& Chen, S.}\ 1998\ The scaling of pressure in isotropic
turbulence. {\it Phys. Fluids} {\bf 10}, 2119-2121.

{\sc Obukhov, A. M., \& Yaglom, A. M.}\ 1951\ The microstructure of
turbulent flow. {\it Prikl. Mat. Mekh.} {\bf 15}, 3-26.

{\sc Pearson, B. R. \& Antonia, R. A.}\ 2001\ Reynolds number dependence of
turbulent velocity and pressure increments. {\it J.~Fluid Mech.} {\bf 444},
343-382.

{\sc Shaw, R. A. \& S. P. Oncley,} 2001 Acceleration intermittency and
enhanced collision kernels in turbulent clouds. {\it Atmos. Res.}\
(accepted).

{\sc Sreenivasan, K. R.} 1995\ On the universality of the Kolmogorov
constant. {\it Phys. Fluids} {\bf 7}, 2778-2784.

{\sc Sreenivasan, K. R. \& Antonia, R. A.} 1997\ The phenomenology of small
scale turbulence. {\it Annu. Rev. Fluid Mech.} {\bf 29}, 435-472.

{\sc Sreenivasan, K. R. \& Kailasnath, P.} 1993\ An update on the
intermittency exponent in turbulence. {\it Phys. Fluids A} {\bf 5}, 512-514.

{\sc Stolovitzky, G., Sreenivasan, K. R. \& Juneja, A.}\ 1993\ Scaling
functions and scaling exponents in turbulence. {\it Phys. Rev. E} {\bf 48},
R3217-R3220.

{\sc Taylor, G. I.} 1935 Statistical theory of turbulence. {\it Proc. Roy.
Soc. London} {\bf 151}, 465-478.

{\sc Tavoularis, S., Bennett, J. C. \& Corrsin, S.} 1978 Velocity-derivative
skewness in small Reynolds number, nearly isotropic turbulence. {\it J.
Fluid Mech.} {\bf 88}, 63-69.

{\sc Vedula, P. \& Yeung, P. K.}\ 1999\ Similarity scaling of acceleration
and pressure statistics in numerical simulations of isotropic turbulence. 
{\it Phys. Fluids} {\bf 11}, 1208-1220.

{\sc Zocchi, G., Tabeling, P., Maurer, J. \& Willaime, H.}\ 1994\
Measurement of the scaling of the dissipation at high Reynolds numbers. {\it %
Phys. Rev. E} {\bf 50}, 3693-3700.

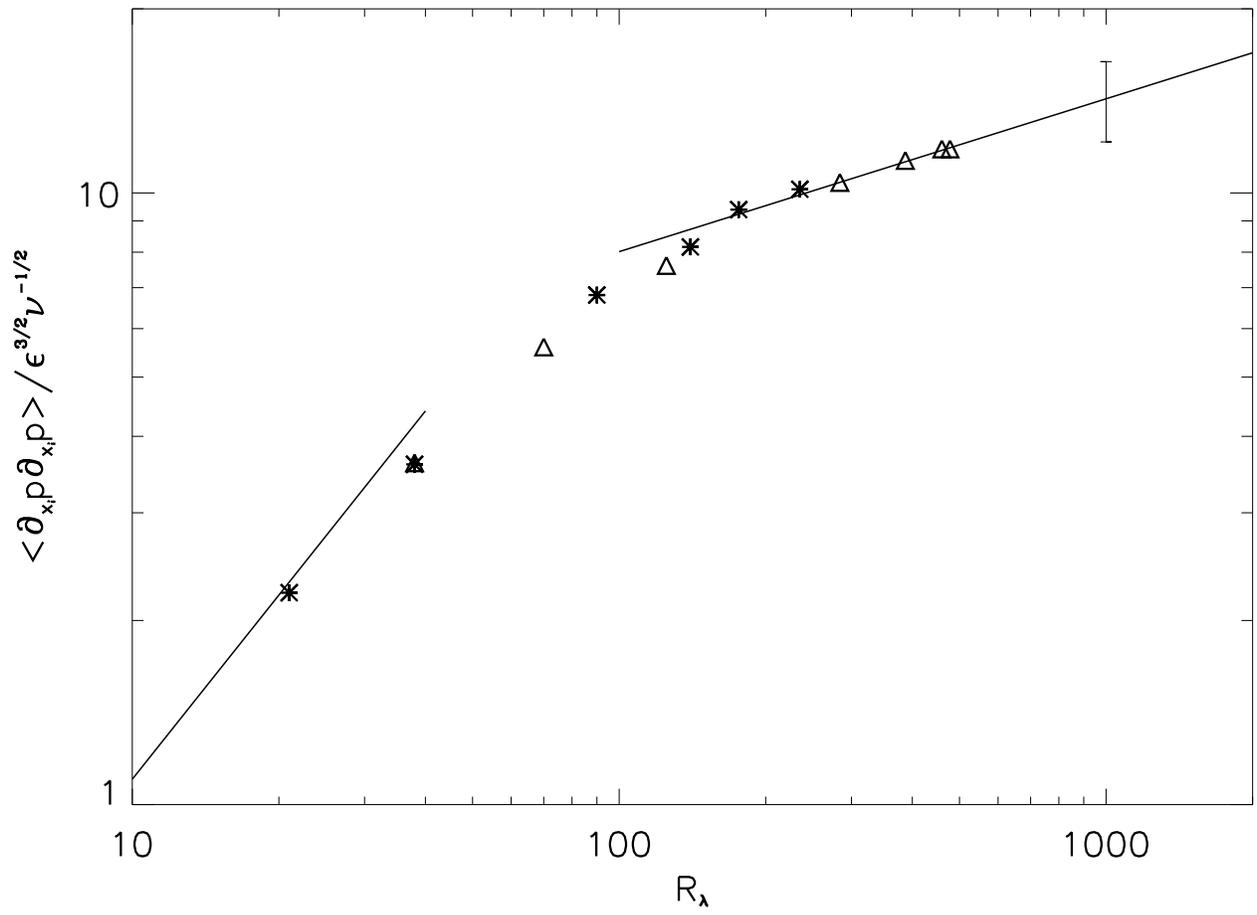
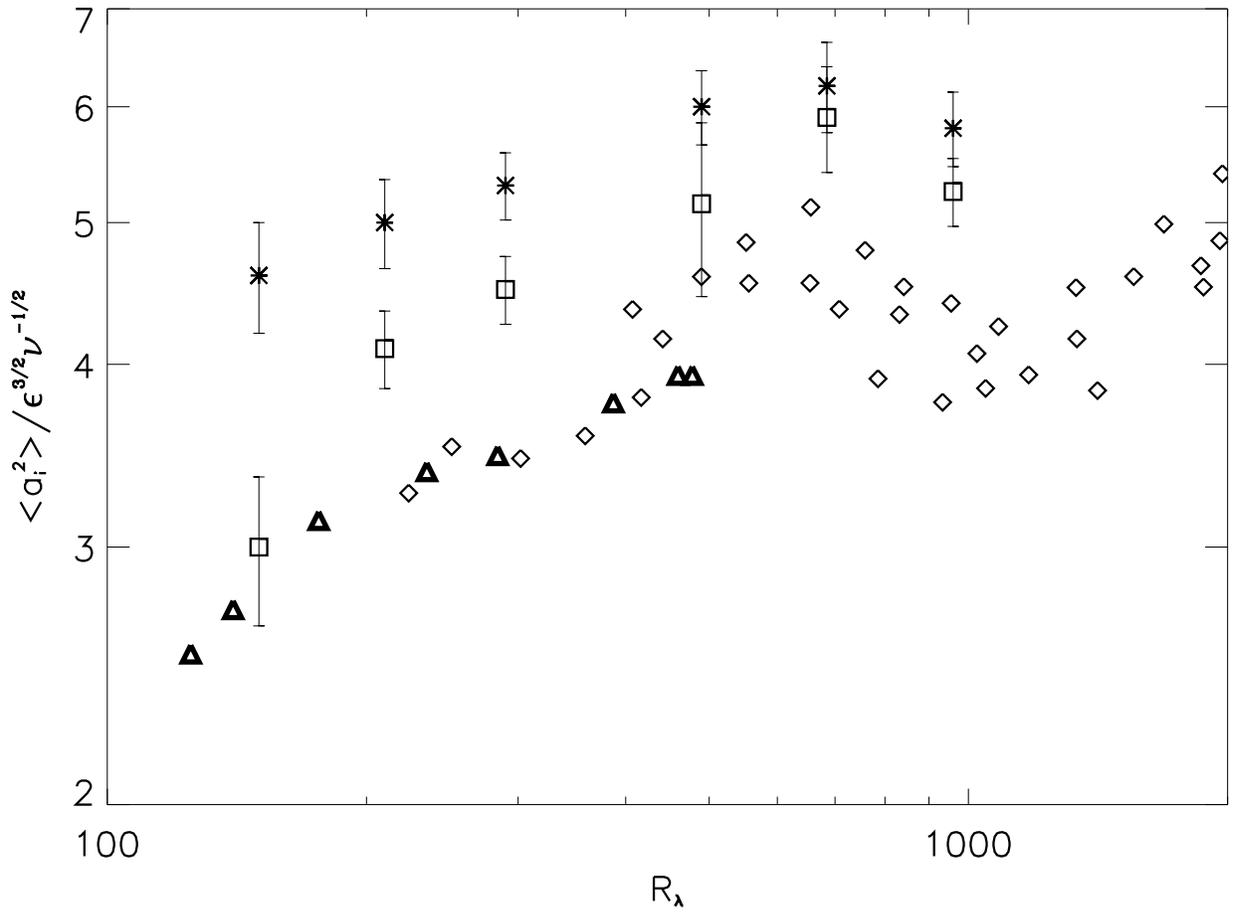
\end{document}